\documentclass[12pt,a4paper]{iopart}
\usepackage{epsf}
\def\la{{\langle}}
\def\ra{{\rangle}}
\def\vep{{\varepsilon}}
\newcommand{\beq}{\begin{equation}}
\newcommand{\eeq}{\end{equation}}
\newcommand{\beqa}{\begin{eqnarray}}
\newcommand{\eeqa}{\end{eqnarray}}

\newcommand{\wh}{\widehat}
\newcommand{\Op}{{\wh{\Omega}_+}}
\newcommand{\Om}{{\wh{\Omega}_-}}
\newcommand{\Opm}{{\wh{\Omega}_{\pm}}}
\newcommand{\Oopm}{{\wh{\Omega}_{+/-}}}
\newcommand{\intf}{\int_{-\infty}^\infty}

\newcommand{\sip}{{\rm{sign}}(p)}

\begin{document}

\title[Step-like potentials]{Moller operators and Lippmann-Schwinger equations
for step-like potentials}
\author{A D Baute\dag\ddag, I L Egusquiza\dag\ and J G Muga\ddag}
\address{\dag\ Fisika Teorikoaren Saila,
Euskal Herriko Unibertsitatea, 644 P.K., 48080 Bilbao, Spain}
\address{\ddag\ Departamento de Qu\'\i mica-F\'\i sica,
Universidad del Pa\'\i s Vasco, Apdo. 644, 48080 Bilbao, Spain}

\begin{abstract}
The Moller operators and the asociated Lippman-Schwinger equations
obtained from different partitionings of the Hamiltonian for a
step-like potential barrier are worked out, compared and related.
\end{abstract}
%
\pacs{03.65.-w\hfill EHU-FT/0102}
\section{Introduction}
One dimensional (1D) quantum scattering theory is usually formulated
for potentials that vanish asymptotically both for large positive and
negative values of the coordinate $x$. It is well known that the
degeneracy of the energy makes the full-line scattering problem
somewhat more involved than partial-wave scattering on the half-line.
Additional complications arise for step-like potentials, namely, when
the potential tends to different constant values on both sides,
\[
\lim_{x\to{-\infty}}V(x)= 0\,,\qquad
\lim_{x\to{\infty}}V(x)= V_0>0\,.
\]
These conditions apply for example to electron collisions between
different metals, in models of time-of-arrival measurement
\cite{AOPRU98,BEM01a}, or in experiments with evanescent waves.  In
some cases it is enough to solve the Schr\"odinger equation
numerically, subject to scattering boundary conditions, in order to
obtain the transmission and reflection amplitudes. There are however
applications where a formal theory of scattering is needed. By
``formal theory'' we mean the network of operators (Moller, $\wh{S}$,
$\wh{T}$, and resolvents), which, together with their generic
properties and relations, are used to describe the collision.  These
applications include the obtention of approximate analytical formulae,
perturbative analysis, inverse scattering methods based on
``two-potential formulae'', kinetic theory, or the study of
characteristic times \cite{JW89}. The work on the scattering theory of
step-like potentials has concentrated on the inverse problem
\cite{BF62,CK85,Legendre82}, characterizations of scattering data for
classes of potentials \cite{CK85}, zero energy limits
\cite{Aktosun99}, Levinson's theorem \cite{Aktosun99}, and compact
formulae for the evolution of states with initial support on one
half-line \cite{BEM01,HWZ77}.  This paper complements those mentioned
by focusing on the formal setting of the theory. In particular, we
stress the fact that several partitionings of the Hamiltonian are
possible, and work out, compare and relate the Moller operators and
the corresponding Lippmann-Schwinger (LS) equations derived from them.
Compact expressions of the asymptotic transmission and reflection
amplitudes are given in terms of different potential-dependent matrix
elements.  The formalism is presented with ``physicist's rigor''. Its
validity is in any case easily checked for cut-off potentials that
deviate from the two asymptotic values $0$ and $V_0$ only in a finite
domain, $[a,b]$, which is the case explicitly considered throughout.
It is expected though that it will apply for other potentials as well,
having smooth but sufficiently rapid decay.

For completeness, and in order to introduce the relevant concepts and
notation, in section II we present a lightnight review of Moller
operators and Lippmann-Schwinger equations for potentials that vanish
on both sides (the ``ordinary case'' hereafter), while some properties
of scattering states of the Hamiltonian $\wh{H}$ for step-like
potentials are to be found in section III. We discuss several
partitionings of the Hamiltonian, together with the corresponding
Moller operators and Lippmann-Schwinger equations in the following
sections IV, V and VI. So as best to illustrate the differences among
the formalisms we address the issue of the existence of Born's
approximation in section VII.

\section{Moller operators for potentials that vanish on both sides}
In ordinary 1D scattering the Moller operators $\Opm$, defined by the
strong limits
\beq\label{opm}
\Opm=\lim_{t\to\mp\infty} e^{i\wh{H}t/\hbar}e^{-i\wh{H}_0 t/\hbar},
\eeq
link the actual state $\psi$ with its asymptotic free-motion reference
states, $\phi_{in}$ and $\phi_{out}$,
\[
\lim_{t\to{-/+\infty}}||\psi(t)-\phi_{in/out}(t)||= 0\,,
\]
The operator $\Op$ (respectively $\Om$) provides the scattering state
by acting on the incoming (resp. outgoing) asymptote, $\phi_{in}$
(resp. $\phi_{out}$) ,
\beq
\label{psiin}\Oopm |\phi_{in/out}(t)\ra=|\psi(t)\ra,
\eeq
for all $t$.

The total Hamiltonian, $\wh{H}=\wh{H}_0+\wh{V}$, is composed by a free
motion Hamiltonian, $\wh{H}_0=\wh{p}^2/2m$, that governs the motion of
the asymptotes, and a potential operator, $\wh{V}$, with a local
coordinate representation $\la x|\wh{V}|x'\ra=\delta(x-x')V(x)$.  The
potential function $V(x)$ vanishes as $|x|\to\infty$, in such a way
that the Moller operators in (\ref{opm}) exist.  For concreteness, we
shall in fact assume that $V(x)$ vanishes outside the finite interval
$[a,b]$, with $a\le 0$ and $b\ge 0$.

The infinite time limits in the definition of $\Opm$,  (\ref{opm}),
may also be expressed with the alternative forms
\[
\Opm=\lim_{\vep\to 0\pm}\mp\vep\int_0^{\mp\infty} dt'\, e^{\pm\vep t'}
e^{i\wh{H}t'/\hbar}e^{-i\wh{H}_0t'/\hbar}\,.
\]
Inserting a resolution of the identity in momenta between $\Opm$ and
$|\phi_{in}(t)\ra$ or $|\phi_{out}(t)\ra$, and integrating over $t'$,
there results
\beq
|\psi(t)\ra=\intf dp\,e^{-iE_pt/\hbar}|p^{+/-}\ra\la
p|\phi_{in/out}(0)\ra,
\label{10}
\eeq
where we have introduced the (improper) eigenstates of $\wh{H}$, with eigenvalue $E_p$,
\beq\label{lsg}
|p^\pm\ra=\wh{\Omega}(E_p\pm i0)|p\ra\equiv|p\ra+\frac{1}{E_p\pm
i0-\wh{H}}
\wh{V}|p\ra,\qquad E_p=p^2/2m. 
\eeq
The states $|p\ra$ (or $|q\ra$, to be used in the following)  are the usual plane wave states, $\la x|p\ra=\exp(ipx/\hbar)/\sqrt{2\pi\hbar}$ (resp. $\la x|q\ra=\exp(iqx/\hbar)/\sqrt{2\pi\hbar}$).
$\wh{\Omega}(z)$ is a {\it parameterized Moller operator} (to be
distinguished from the abstract ones in  (\ref{opm})) which, unlike
$\wh{\Omega}_\pm$, can be applied to plane waves, and can be defined
through
\[
\wh{\Omega}(z)=1+\wh{G}_0(z)\wh{T}(z).
\]
In this equation,
\beq\label{t}
\wh{T}(z)=\wh{V}+\wh{V}\wh{G}(z)\wh{V}\eeq
is the parameterized ``$T$-operator'', or transition $T$ operator, and the operators
\[
\wh{G}(z)\equiv(z-\wh{H})^{-1}\,,{\rm and}
\qquad\wh{G}_0(z)\equiv(z-\wh{H}_0)^{-1}
\]
are the resolvents for the Hamiltonians $\wh{H}$ and $\wh{H}_0$
respectively. Equation (\ref{t}) is called the operator
Lippmann-Schwinger equation.  Expressions equivalent to  (\ref{lsg})
are obtained by using the operator Lippmann-Schwinger equation
(\ref{t}) and the operator identity
\[
\wh{G}_0(z)\wh{T}(z)=\wh{G}(z)\wh{V}\,,\]
which lead to 
\beq\label{lis}|p^{\pm}\ra=|p\ra+\frac{1}{E_p\pm i0-\wh{H}_0}
\wh{T}(E_p\pm i0)|p\ra=|p\ra+\frac{1}{E_p\pm i0-\wh{H}_0}\wh{V}|p^{\pm}\ra.
\eeq
Equations (\ref{lsg}) and (\ref{lis}) are different alternative forms
of the Lippmann-Schwinger integral equation for the states
$|p^{\pm}\ra$. Note the structure of these states, composed by a free
plane wave (incoming for $|p^+\ra$, outgoing for $|p^-\ra$) and a
scattering part.  The forms (\ref{lis}) are useful to determine the
asymptotic behaviour of the states at large distances (for cut-off
potentials this means $x<a$, $x>b$) since the matrix elements of
$G_0(E_p\pm i0)$ (the Green's function) are known,
\[
\langle x|\frac{1}{z 
-\wh{H}_0}|x'\rangle=-\frac{im}{\hbar(2mz)^{1/2}}
\,e^{i(2mz)^{1/2}|x-x'|/\hbar}.
\]
In this expression the square root is defined with a branch cut along
the positive axis.  Using delta-function normalization (i.e., $\langle
p^\pm|{p'}^\pm\rangle=\delta(p-p')$), the states behave outside $[a,b]$ as
\begin{equation}\label{pl}
\la x|p^{{{\rm sign}}(p)}\ra=
\frac{1}{h^{1/2}}\times\cases{\exp(ipx/\hbar)
+R^l(p)\exp(-ipx/\hbar),& $\,\, x<a$,\cr T^l(p)\exp(ipx/\hbar),& $\,\,
x>b$,\cr}
\end{equation}
\begin{equation}\label{pr}
\la x|p^{{-{\rm{sign}}(p)}}\ra=
\frac{1}{h^{1/2}}\times
\cases{T^r(-p)\exp(ipx/\hbar),& $\,\, x<a$,\cr
\exp(ipx/\hbar)+R^r(-p)\exp(-ipx/\hbar),& $\,\, x>b$.\cr}
\end{equation}
Both in (\ref{pl}) and (\ref{pr}) $p$ is a label for the energy. Let
us first interpret the states in (\ref{pl}): for $p>0$, there is an
incident plane wave from the left, with wavenumber $p/\hbar$, and
$R^l(p)$ and $T^l(p)$ are the corresponding reflection and
transmission amplitudes for left incidence; on the other hand, if
$p<0$, there is an outgoing plane wave towards the left, with
wavenumber $|p|/\hbar$, and the corresponding amplitudes are not
properly related to ``transmission" and ``reflection". However, since
they are analytical continuations of the amplitudes for $p>0$, the
same notation is mantained.  Similar considerations apply to the set
of states described by  (\ref{pr}).

The particular form of the amplitudes $T^{l,r}(p)$ and $R^{l,r}(p)$
for potentials composed by square barriers is easily obtained by
matching the wave function and its derivative at the edges. However,
this procedure is useless in more general cases. Expressions of the
amplitudes for the general case are obtained by comparing
 (\ref{pl}) and (\ref{pr}) with the coordinate representation of
 (\ref{lis}).  In this way they can be related to
on-the-energy-shell elements of the transition operators
$\wh{T}(E_p\pm i0)$.  We shall work out one case in detail, as a
reference for later results.  Assume $p>0$ and $x>b$. In
\[
\la x|p^+\ra=\la x|p\ra+\intf dx' \la x|\wh{G}_0(E_p+i0)|x'\ra\la x'|
\wh{T}(E_p+i0)|p\ra,
\]
we can substitute $|x-x'|$ in the Green's function by $x-x'$, since
the support of $\la x'| \wh{T}(E_p+i0)|p\ra$ is necessarily restricted
to be between $a$ and $b$ because of the dependence of $\wh{T}$ on
$\wh{V}$, see (\ref{t}), and the finite support of $V(x)$.  Therefore,
\beqa
\la x|p^+\ra&=&\la x|p\ra 
-\frac{2\pi m i}{h} \frac{e^{ipx/\hbar}}{p}\intf dx'\,e^{-ipx'/\hbar}
\la x'|\wh{T}
(E_p+i0)|p\ra \nonumber\\
&=&\la x|p\ra-\frac{2\pi m i}{p} \la x|p\ra T_{p,p}^+\,,\nonumber
\eeqa
where
\[
T^\pm_{p,p'}\equiv\la p|\wh{T}(E_p\pm i0)|p'\ra,\;\; |p|=|p'|.
\]
Straightforward comparison with (\ref{pl}) leads to an explicit
expression for $T^l(p)$.  The rest of the amplitudes can be worked out
similarly to obtain the following table
\beqa\nonumber
T(p)&=&1-\frac{2i\pi m}{p}T_{p,p}^{{\rm sign}(p)},
%
%
\\
R^l(p)&=&-\frac{2mi\pi}{p}T_{-p,p}^{{\rm sign}(p)},\label{rlp}
\\
R^r(p)&=&-\frac{2mi\pi}{p} T_{p,-p}^{{\rm sign}(p)}.\nonumber
\eeqa
Note that time reversal invariance implies
$T_{p,p}^{\pm}=T_{-p,-p}^\pm$, and therefore $T^r(p)=T^l(p)=T(p)$.
\section{Scattering eigenstates of the Hamiltonian for step-like potentials}
In the case of step-like potentials, the potential function $V(x)$
does not go to zero both for positive and negative $x$, when
$|x|\to\infty$. We shall assume in what follows that $V(x)$ does
indeed tend to zero as $x\to-\infty$, and to $V_0$ when $x\to+\infty$.
In other words, we shall assume that $V(x)$ equals $V_\theta(x)=V_0 \theta(x)$
plus some localized additional potential of finite support or that
tends to zero sufficiently fast when $|x|\to\infty$. In such a case,
the scattering part of the energy spectrum is doubly degenerate above $V_0$, as corresponds
physically to incidence from one side or the other.  Below $V_0$,
however, there is only one linearly independent solution with an
evanescent wave at $x>0$.  There may be bound states too, with energy
$E_j<0$.  The resolution of the identity may be written in different
ways, in particular as \cite{BEM01}
\[
\wh{1}=\sum_j|E_j\rangle \langle E_j| +\int_{-\infty}^{-p_0}dp\,
|p^\pm\rangle\langle p^\pm| +\int_{p_{0}}^\infty
dp\,|p^\pm\rangle\langle p^\pm|
\pm\int_{0}^{\pm p_{0}}dp\,|p^\pm\rangle\langle p^\pm|,
\]
where $p_0=(2mV_0)^{1/2}$ and the states $|p^\pm\rangle$, to be
defined below, have as in the ordinary case an energy
$E_p=p^2/(2m)$. As pointed out above, $p$ is 
 a label of the energy. It can be positive or negative because of the degeneracy in energy.

The states $|p^+\rangle$, with $p>0$, have an {\it incident} plane
wave of wavenumber $p/\hbar$, and the states $|p^-\ra$, $p<0$, a
corresponding outgoing one,
\begin{equation}\label{plstep}
\la x|p^{{\rm sign}(p)}\ra=
\frac{1}{h^{1/2}}\times\cases{\exp(ipx/\hbar)
+R^l(p)\exp(-ipx/\hbar),& $\,\, x<a$\cr T^l(p)\exp(iqx/\hbar),& $\,\,
x>b$\cr},
\end{equation}
where $q=(p^2-2mV_0)^{1/2}$, with a branch cut that joins the branch
points $p=\pm p_0$, going slightly below ${\rm Im}(p)=0$.  In this way
the sign of $q$ is the same as the sign of $p$ for $p^2>p_{0}^2$,
whereas it becomes positive imaginary for $-p_0<p<p_0$.

The states $|p^+\rangle$ for $p<-p_0$ are defined by an {\it incident}
plane wave from the right with wavelength $-\hbar/q(>0)$, and states
$|p^-\ra$ with $p>p_0$ by an outgoing plane wave with wavelength
$\hbar/q$,
\begin{equation}\label{prstep}
\la x|p^{-{\rm sign}(p)}\ra
=\frac{1}{h^{1/2}}\left(\frac{p}{q}\right)^{1/2}\times
\cases{T^r(-p)\exp(ipx/\hbar),& $\, x<a$\cr
\exp(iqx/\hbar)+R^r(-p)\exp(-iqx/\hbar),& $\, x>b$\cr},
\end{equation}
(always for $|p|>|p_0|$).
The factor $(p/q)^{1/2}$ is necessary for the proper delta
normalization, that is, $\langle p^+|{p'}^+\rangle=\delta(p-p')$, and
the corresponding expression for the $|p^-\ra$ scattering states.  As
in the ordinary case, the arguments of transmission or reflection
amplitudes are always positive for states $|p^+\rangle$, and negative
for states $|p^-\ra$ independently of the sign of $p$.
%
 
The ${\sf S}$ matrix elements are defined as the coefficients
multiplying the outgoing plane waves when the incident plane wave is
normalized to unit flux.  When both channels are open ($p>p_0$), the
${\sf S}$ matrix reads
\[
{\sf S}(p)=\left(
\matrix{
\left(\frac{q}{p}\right)^{1/2}T^l(p) & R^l(p)
\cr
R^r(p) & \left(\frac{p}{q}\right)^{1/2}T^r(p)
\cr} 
\right)\,.
\]
One may also obtain these matrix elements from $\la p'^-|p^+\ra$ by
factoring out a delta function in the scattering energy.  The unitarity of the
${\sf S}$ matrix, ${\sf S}{\sf S}^\dagger=1$, implies relations among
the amplitudes,
\beqa
\frac{p}{q}|T^r(p)|^2+|R^r(p)|^2=1\,,\nonumber
\\
\label{uni1}
|R^l(p)|^2+\frac{q}{p}|T^l(p)|=1\,,
\\
\frac{p}{q}T^r(p)R^l(p)^*+R^r(p)T^l(p)^*=0\,.\nonumber
\eeqa
For $0<p<p_0$ only one channel is open, the ${\sf S}$ matrix reduces
to a number, $R^l(p)$, and unitarity implies
\beq
\label{uni4}
R^l(p)R^l(p)^*=1.
\eeq
All these equations,  the set (\ref{uni1}) and  (\ref{uni4}), are also valid
for negative label $p$, thus providing relations for the amplitudes
associated with $|p^-\ra$ states.
\section{Step-like potentials. Multichannel formalism.}
The straightforward application of the Moller operators of section II,
based on the partitioning $\wh{H}=\wh{H}_0+\wh{V}$, to step-like
potentials is justified physically only for certain states.  The key
point is that $\wh{H}_0$ by itself only governs the asymptotic states
that enter from the left (with incident positive momentum), or escape
to the left (with negative outgoing momentum).  So the
Lippmann-Schwinger equations presented in the previous section (that
is, eqns. (\ref{lsg}) and (\ref{lis})), will only be applicable for
$\{|p^{\sip}\ra\}$.  It will prove useful to rename $\wh{H}_0$ as
$\wh{H}_l\equiv \wh{H}_0$, since it is the Hamiltonian associated with
the ``left'' asymptotic channel.  Correspondingly we define
$\wh{V}_l\equiv \wh{V}$, so that the total Hamiltonian is partitioned
as $\wh{H}=\wh{H}_l+\wh{V}_l$, and $\Opm^l\equiv\Opm$.  Similarly, the
states $|p^{-\sip}\ra$, $|p|>|p_0|$, ``start'' (for $p<-p_0$) or ``end
up'' ($p>p_0$) in the right, where the asymptotic Hamiltonian is
$\wh{H}_r\equiv\wh{H}_0+V_0$.  We thus define $\wh{V}_r\equiv
\wh{V}-V_0$, so that $\wh{H}=\wh{H}_r+\wh{V}_r$, and the corresponding
Moller operators
\[
\Opm^r\equiv\lim_{t\to\mp\infty} e^{i\wh{H}t/\hbar}e^{-i\wh{H}_rt/\hbar}.
\]
The asymptotic Hamiltonians have their own resolvents,
\[
\wh{G}_\alpha(z)\equiv\frac{1}{z-\wh{H}_\alpha},
\]
where $\alpha=r,l$ is the subscript to indicate the channel. Notice
that $\wh{G}_l(z)=\wh{G}_0(z)$, using the notation of section II, whereas
$\wh{G}_r(z)=\wh{G}_0(z-V_0)$.  Using the abstract Moller operators one may
define parameterized ones, the corresponding LS equations thus taking
the form
\beqa
|p^{\sip}\ra&=&|p\ra+\wh{G}_l[E_p+\sip i0]\wh{V}_l|p^{\sip}\ra=\nonumber\\
&=&|p\ra+\wh{G}[E_p+\sip i0]\wh{V}_l|p\ra,\label{ppos}
\\
|p^{-\sip}\ra&=&|q_N\ra+\wh{G}_r[E_p-\sip i0]\wh{V}_r|p^{-\sip}\ra=\nonumber\\
&=&|q_N\ra+\wh{G}[E_p-\sip i0]\wh{V}_r|q_N\ra,\;\;\;|p|>|p_0|,\label{pneg}
\eeqa
where $\la x|q_N\ra=(p/hq)^{1/2}\exp(ixq/\hbar)$.  A noticeable
difference with the ordinary case is that now the potential functions
$V_\alpha(x)$ are not localized ($V_l(x)$ and $V_r(x)$ do not vanish
for $x>b$ and $x<a$ respectively), so that the simple manipulations
leading, for example, to (\ref{rlp}), are not valid any more to obtain
expressions for $T^l$ and $T^r$. We cannot separate the exponential
$e^{ip|x-x'|/\hbar}$ into $x$ and $x'$ dependent exponentials, and
extract right away the $x$ dependence.  The separation can be done
however to obtain $R^l$ and $R^r$, which take the form
\beqa\label{rl}
R^l(p)&=&\frac{-2\pi im}{p}\la -p|\wh{V}_l|p^{{\rm sign}(p)}\ra,
\\
\label{rr}
R^r(p)&=&\frac{-2\pi im}{p}\la q_N|\wh{V}_r|-p^{{\rm sign}(p)}\ra,
\eeqa
To obtain expressions for the transition amplitudes we rewrite the LS
equations in terms of the potential of the other channel, see Appendix
A,
\beqa
\label{ppos2}
|p^{\sip}\ra&=&\wh{G}_r(E_p+{\sip}0)\wh{V}_r|p^{\sip}\ra,
\\
\label{pneg2}
|p^{-\sip}\ra&=&\wh{G}_l(E_p-{\sip}0)\wh{V}_l|p^{-\sip}\ra\;\;\;\;|p|>|p_0|.
\eeqa
Since the potentials in (\ref{ppos2}) and (\ref{pneg2}) vanish in
regions of space different from the ones in (\ref{ppos}) and
(\ref{pneg}), we may now find the missing expressions for the
transmission amplitudes,
\beqa
\label{tl}
T^l(p)&=&\frac{-2\pi im}{q}\la q|\wh{V}_r|p^{{\rm sign}(p)}\ra,
\\
\label{tr}
T^r(p)&=&\frac{-2\pi im}{p}\la -p|\wh{V}_l|-p^{{\rm sign}(p)}\ra.
\eeqa

\section{Jaworski-Wardlaw Moller operators}

In their study of the time spent by a quantum particle in a given
spatial region \cite{JW89}, Jaworski and Wardlaw introduced two
different asymptotic Hamiltonians for incoming and outgoing
asymptotes,
\beqa
\wh{H}_{in}&=&\frac{\wh{p}^2}{2m}+V_0 \wh{F}_-\nonumber
\\
\wh{H}_{out}&=&\frac{\wh{p}^2}{2m}+V_0 \wh{F}_+\nonumber
\eeqa
where $\wh{F}_-$ and $\wh{F}_+$ are complementary projectors,
$\wh{F}_-+\wh{F}_+=\wh{1}$, over negative and positive momenta
respectively,
\[
\wh{F}_\pm=\pm\int_0^{\pm\infty} dp\, |p\ra\la p|.
\]
Correspondingly, they defined Moller operators
\beq\label{opmjw}
\Oopm^{JW}=\lim_{t\to -/+\infty} e^{i\wh{H}t/\hbar}e^{-i\wh{H}_{in/out}/\hbar}.
\eeq
Note that, as in the previous section, two different partitionings of
the Hamiltonians are required, one for each Moller operator. They are
however not based on right/left channels, but on a distinction between
incoming and outgoing states.  The physical reason for these
definitions is clear: the positive momentum part of the incoming
asymptotes travels on the lower level at long negative times, whereas
the negative momentum parts travels on the upper level. The outgoing
asymptotes behave in the opposite way, with positive momenta on the
upper level and negative momenta on the lower level at large positive
times.

We shall now extend this formalism to produce the asociated
Lippmann-Schwinger equations.  First it is convenient to introduce a
delta-normalized eigenbasis for $\wh{H}_{in}$ and $\wh{H}_{out}$,
(explicitly, $\la in(p)|in(p')\ra=\delta(p-p')$, and similarly for $| out(p)\ra$)
\beqa
\la x|in(p)\ra&=&h^{-1/2}\times\cases{e^{ipx/\hbar}&$p>0$\cr
|p/q|^{1/2}e^{iqx/\hbar}&$p<-p_0$\cr}\nonumber
\\
\la x|out(p)\ra&=&h^{-1/2}\times\cases{|p/q|^{1/2}e^{iqx/\hbar}&$p>p_0$\cr
e^{ipx/\hbar} &$p<0$\cr}\nonumber
\eeqa
so that
\[
\wh{H}_{in/out}|in/out(p)\ra=E_p|in/out(p)\ra.
\]
Aside from the ordinary (momentum) resolution of the identity,
$\wh{1}=\intf dp\,|p\ra\la p|,$
\beqa
\wh{1}&=&\int_{-\infty}^{-p_0}dp\,|in(p)\ra\la in(p)|
+\int_0^\infty dp\,|in(p)\ra\la in(p)|
=\nonumber\\
&=&\int_{-\infty}^{0}dp\,|out(p)\ra\la out(p)| +\int_{p_0}^\infty
dp\,|out(p)\ra\la out(p)|.\nonumber
\eeqa
The connection between the abstract Moller operators (\ref{opmjw}) and
Lippmann-Schwinger equations for eigenstates of $\wh{H}$ follows now
closely the steps from (\ref{psiin}) to (\ref{lis}), but making use of
the above resolutions of the identity.  We thus find
\beqa
|\psi(t)\ra&=&\int_{-\infty}^{-p_0}dp\,|p^+\ra\la
in(p)|\phi_{in}(t)\ra +\int_0^\infty dp\,|p^+\ra\la
in(p)|\phi_{in}(t)\ra\,,\nonumber
\\
|\psi(t)\ra&=&\int_{-\infty}^{0}dp\,|p^-\ra\la out(p)|\phi_{out}(t)\ra
+\int_{p_0}^\infty dp\,|p^-\ra\la out(p)|\phi_{out}(t)\ra\,,\nonumber
\eeqa
with
\beqa
\label{JW+}
|p^+\ra&=&|in(p)\ra+\wh{G}_{in}(E_p+i0)\wh{T}_{in}(E_p+i0)|in(p)\ra,
\\ 
\label{JW-}  
|p^-\ra&=&|out(p)\ra+\wh{G}_{out}(E_p-i0)\wh{T}_{out}(E_p-i0)|out(p)\ra,
\eeqa
and
\beqa
\wh{G}_{in/out}(z)&=&(z-\wh{H}_{in/out})^{-1},
\\
\wh{T}_{in/out}(z)&=&\wh{V}_{in/out}+\wh{V}_{in/out}
\wh{G}(z)\wh{V}_{in/out},\label{lipsdos}
\eeqa
corresponding to the two partitionings of the Hamiltonian,
$\wh{H}=\wh{H}_{in}+\wh{V}_{in}=\wh{H}_{out}+\wh{V}_{out}$, where
\[
\wh{V}_{in}=\wh{V}-V_0 \wh{F}_-\,,\quad{\rm and}\quad
\wh{V}_{out}=\wh{V}-V_0 \wh{F}_+\,.
\]
However, the potentials $\wh{V}_{in}$ and $\wh{V}_{out}$ are not
localized. They do not vanish as $x\to\infty$ and this leads to
similar problems to the ones encountered before when searching for
expressions for the transmission amplitudes.  They are actually more
severe now because these potentials do not have a semibounded support;
in addition, the zeroth order Green's functions, which can be
explicitly obtained by integration in the complex momentum plane, are
cumbersome to work with,
\beqa
\label{gin}
\la x|\wh{G}_{in}(z)|x'\ra&=&\la x |\frac{\wh{F}_+}{z-\wh{H}_0}|x'\ra+
\la x|\frac{\wh{F}_-}{z-V_0-\wh{H}_0}|x'\ra,
\\
\la x|\wh{G}_{out}(z)|x'\ra&=&\la x |\frac{\wh{F}_+}{z-V_0-\wh{H}_0}|x'\ra+
\la x|\frac{\wh{F}_-}{z-\wh{H}_0}|x'\ra. 
\eeqa
The summands are particular cases of
\[
\la x|\frac{\wh{F}_\xi}{\zeta -\wh{H}_0}|x'\ra=
A\xi{\rm sign}(x-x')+\theta[\xi (x-x')]
\la x|\wh{G}_0(\zeta)|x'\ra,
\]
with $\xi=\pm $ and
\beqa
\nonumber
A&=&\frac{2mi}{h(2m\zeta )^{1/2}}[{\rm ci}(y)\sin(y)-{\rm si}(y)\cos(y)],
\\
y&=&(2m\zeta )^{1/2}|x-x'|/\hbar,
\label{a}
\eeqa
where the square root with positive imaginary part is taken.

The scattering states defined through  (\ref{JW+}) and (\ref{JW-})
are the same as those defined previously by the LS equations of the
multichannel method presented in the previous section. In order to
check the veracity of this statement, it is convenient to use the
identity $\wh{G}_{in/out}(z)\wh{T}_{in/out}(z)
=\wh{G}(z)\wh{V}_{in/out}$, together with the forms of the resolvents
$\wh{G}$ given in  (\ref{ppos}) and (\ref{pneg}), and the defining
expressions of the different potential operators involved.

\section{Pure-step Hamiltonian as zeroth order}  
In this section we shall study one more possible partitioning of the
Hamiltonian, by considering the Hamiltonian $\wh{H}_s=
\wh{H}_0+\wh{V}_\theta$ for the pure step potential
$\wh{V}_\theta\equiv V_0\theta(\wh{x})$, as the zeroth order term for
the complete Hamiltonian,
\[
\wh{H}=\wh{H}_{s}+\wh{V}_s.
\]
In other words, the total potential energy is decomposed into the pure
step potential part and a localized part, as
$\wh{V}=\wh{V}_\theta+\wh{V}_s$.
It is easy to compute two different eigenbases of $\wh{H}_s$, whose
elements are $|p^\pm_s\ra$ respectively (labeled by $p$ as before).
Their explicit expression lends itself to identification of
transmission and reflection amplitudes by comparison with expressions
(\ref{plstep}) and (\ref{prstep}):
\beqa
T^l_{s}(p)&=&\frac{2p}{q+p},\;\;\;\;\;\;R^l_{s}(p)=\frac{p-q}{q+p},
\nonumber\\
T^r_{s}(p)&=&\frac{2q}{p+q},\;\;\;\;\;\;R^r_{s}(p)=\frac{q-p}{p+q}.
\label{step}
\eeqa
Green's function for $\wh{H}_s$ is also known exactly
\cite{Aguiar93},
\beq
\label{Gs}
\la x|\wh{G}_s(E_p\pm i0)|x'\ra=
\pm\frac{m}{i\hbar}
\cases{
\frac{1}{|p|}\big[e^{\pm i|p||x-x'|/\hbar}
+r_\pm e^{\mp i|p|(x+x')/\hbar} \big],
&$x'<0,\;x<0$
\cr      
\frac{1}{|p|}t_\pm e^{\pm i(\mu_\pm x-|p|x')/\hbar},&$x'<0,\; x>0$
\cr
\frac{1}{|p|}t_\pm e^{\pm i(\mu_\pm x'-|p|x)/\hbar},&$x'>0,\; x<0$
\cr             
\frac{1}{\mu_\pm}\big[e^{\pm i\mu_\pm |x-x'|/\hbar}
-r_\pm e^{\pm i\mu_\pm(x+x')/\hbar}\big],&$x'>0,\; x>0$
\cr}
\eeq
where $E_p=|p|^2/2m$,
\beqa
t_\pm&=&\frac{2|p|}{|p|+\mu_\pm},\nonumber
\\
r_\pm&=&\frac{|p|-\mu_\pm}{|p|+\mu_\pm},\nonumber
\eeqa 
and
\[
\mu_\pm=\cases{
[2m(E_p-V_0)]^{1/2},& $E_p>V_0$
\cr
\pm i[2m(V_0-E_p)]^{1/2},& $E_p<V_0$
\cr}.   
\]
A first advantage of this decomposition is that the state is governed
asymptotically by $\wh{H}_s$ both before and after the collision, to
the right and to the left, so that the physically meaningful Moller
operators can be defined, as in the ordinary case, by the two limits
of a unique operator expression,
\[
\Opm^s=\lim_{t\to\mp\infty}e^{i\wh{H}t/\hbar}e^{-i\wh{H}_{s}t/\hbar}, 
\]
which amounts to a formal simplification with respect to the
partitionings of the two previous sections, and absence of extra
indices.  Analogous steps to those leading to (\ref{10}), with the
decomposition of unity in the basis of $\wh{H}_{s}$, provide us with
\beqa
|\psi(t)\ra&=&\int_{-\infty}^{-p_0} dp\, |p^+\ra \la p^+_s|\phi_{in}(t)\ra
+\int_{0}^{\infty} dp\, |p^+\ra \la p^+_s|\phi_{in}(t)\ra\,,\nonumber
\\
|\psi(t)\ra&=&\int_{-\infty}^{0} dp\, |p^-\ra \la p^-_s|\phi_{out}(t)\ra
+\int_{p_0}^{\infty}dp\, |p^-\ra \la p^-_s|\phi_{out}(t)\ra\nonumber
\eeqa
where
\beq\label{lss}
|p^\pm\ra=|p^\pm_s\ra+\wh{G}_s(E_p\pm i0)
\wh{T}_s(E_p\pm i0)|p^\pm_s\ra
\eeq
and
\[
\wh{T}_s(z)=\wh{V}_s+\wh{V}_s\wh{G}(z)\wh{V}_s\,.
\]
A second advantage of this decomposition is that the potential
function $V_s(x)=V(x)-V_0\theta(x)$ is now localized.  One may thus
obtain easily the explicit $x$-dependence of the coordinate
representation of (\ref{lss}) and identify expressions for the
transmission and reflection amplitudes in terms of the localized
potential,
\beqa
T^l(p)&=&T^l_{s}(p)-\frac{2\pi mi}{q}\left(\frac{q}{p}\right)^{1/2}
\la p_s^{-{\rm sign}(p)}|\wh{T}_s[E_p+ {\rm sign}(p)
i0]|p_s^{{\rm sign}(p)}\ra\,,\nonumber
\\
T^r(p)&=&T^r_{s}(p)-\frac{2\pi mi}{p}\left(\frac{q}{p}\right)^{1/2}
\la -p_s^{-{\rm sign}(p)}|\wh{T}_s[E_p+ {\rm sign}(p) i0]
|-p_s^{{\rm sign}(p)}\ra\,,\nonumber
\\
\label{rls}
R^l(p)&=&R^l_{s}(p)-\frac{2\pi mi}{p}
\la -p_s^{-{\rm sign}(p)}|\wh{T}_s[E_p+ {\rm sign}(p) i0]
|p_s^{{\rm sign}(p)}\ra
\\
R^r(p)&=&R^r_{s}(p)-\frac{2\pi mi}{p}
\la p_s^{-{\rm sign}(p)}|\wh{T}_s[E_p+ {\rm sign}(p) i0]
|-p_s^{{\rm sign}(p)}\ra\,,\nonumber
\eeqa
The time-reversal antiunitary operator $\Theta$ changes the sign of
$\Opm^{s}$, $\Theta\Opm^{s}=\wh{\Omega}_\mp\Theta$, as in the ordinary
case.
From the time reversal invariance of the Hamiltonian it follows that
$\la p|\wh{T}_s[E_p+\sip i0]|p'\ra=\la -p|\wh{T}_s[E_p+\sip
i0]|-p'\ra$ (on the energy shell), so that the transmission amplitudes
are related by $T^r(p)=(q/p)T^l(p)$.

The agreement with the previous compact expressions
(\ref{rl}-\ref{rr}), and (\ref{tl}-\ref{tr}) is found by using
(\ref{rl}-\ref{rr}) and (\ref{tl}-\ref{tr}) themselves for the step
potential $\wh{V}_\theta$, and the following non trivial
generalizations of the standard ``two-potential'' formula to the two
partitionings of the multichannel formalism (see Appendix B),
\beqa\nonumber
\la p|\wh{V}|\pm p^{-{\rm sign}(p)}\ra&=&
\la p|\wh{V}_\theta|\pm p_s^{-{\rm sign}(p)}\ra
+\la p_s^{{\rm sign}(p)}|\wh{V}_s|\pm p^{-{\rm sign}(p)}\ra,
\\
\nonumber
\la q_N|(\wh{V}-V_0)|\pm p^{{\rm sign}(p)}\ra&=&
\la q_N|(\wh{V}_\theta-V_0)|\pm p_s^{{\rm sign}(p)}\ra
+\la p_s^{-{\rm sign}(p)}|\wh{V}_s|\pm p^{{\rm sign}(p)}\ra.
\eeqa

The use of the bra-ket notation, while standard and very convenient
most of the time, requires some greater attention than usual to describe
adequately the evanescent case, when $q=i\gamma$,
$\gamma>0$. Irrespective of the value of $p$, $\la q|x\ra$ should
always be interpreted as $h^{-1/2}\exp(-iqx)$.  Similarly, $\la
p^\pm_s|x\ra$ should {\it first} be written for real $q$ and {\it
then} continued analytically.
\section{Born approximations}
As an example to illustrate the differences of the three described
formalisms we shall obtain the Born approximation of the reflectance
$|R^l(p)|^2$ for the potential,
\beq\label{pote}
\wh{V}=\wh{V}_\theta+V_1\delta(\wh{x}).
\eeq
The exact result,
\[
R^l(p)=
\frac{p-q-(2imV_{1}/\hbar)}{p+q+(2imV_{1}/\hbar)},
\]
may be obtained using (\ref{step}), (\ref{rls}),
$\delta(\wh{x})=|0\ra\la 0|$, and
\[
\wh{T}_s(z)=\frac{V_{1}|0\ra\la 0|}{1-V_{1}\la 0|\wh{G}_{s}
(z)|0\ra},
\]
or alternatively by straightforward computation.

We will now calculate the {\it different} Born approximations by
retaining only the terms linear in the potential corresponding to each
partitioning of the Hamiltonian.  To be more precise, we will look at
the Lippmann-Schwinger equation for $|p^+\ra$ in each approach, and
retain terms of first order in the potential, that is, to first order
in the difference between the total hamiltonian and the incoming
asymptotic hamiltonian of reference. The resulting wavevector will be
examined in the position representation for $x<0$, and the result
compared to  (\ref{plstep}) to extract $R^l(p)$. In fact this last
step is not necessary for the multichannel (MM - sec. IV) and the
localized potential (LP - sec. VI) approaches, since we have already
carried out this comparison in an exact manner (see  (\ref{rl})
and (\ref{rls})).  Notice that we have indeed checked that in all
three approaches we obtain the same scattering state $|p^+\ra$.

In the multichannel method (MM), see (\ref{rl}), the first order term
in $\wh{V}_l$ (which in this case is $\wh{V}_l=\wh{V}_\theta+V_1\delta(\wh{x})$) is
\beq
R^l_{Born-MM}(p)=\frac{-2\pi mi}{p}\la
-p|\wh{V}_\theta+V_1\delta(\wh{x})|p\ra=
\frac{m(V_0-2ipV_1/\hbar)}{2p^2}.\label{bornmm}
\eeq
The analysis to second order is more delicate, involving limits (as in
$\wh{G}(E_p+i0)$), but it reveals that the singularity in $p=0$ for
the MM formalism actually worsens (it becomes of the form
$p^{-4}$). This was only to be expected, given the non locality of the
perturbing potential in that case, which produces infrared
singularities to all perturbation orders, which can only be resolved
by a complete resummation of all terms.

We could also examine $R^r(p)$, to first order in $\wh{V}_r=\wh{V}_\theta+V_1\delta(\wh{x})-V_0$, which in this case can be obtained
from $R^l(p)$ by substituting $p$ for $q$, and viceversa. This recipe
actually holds for the Born approximation in the MM formalism, which
leads to the result that the reflectance diverges for $|p|\to p_0$.  

A Born approximation in the ``in/out'' formalism of section V is much more
problematic: for $x<0$ and $p>0$, the first order in
$\wh{V}_{in}$ of (\ref{JW+}) is
\[
\la x|p^{+}\ra^{(1)} = \la x|in(p)\ra
+\intf dx'\,\la x|\wh{G}_{in}(E_p+i0)|x'\ra
\la x'|\wh{V}_{in}|in(p)\ra. 
\]
By substituting $\la x|\wh{G}_{in}(E_p+i0)|x'\ra$, see (\ref{gin}) and
({\ref{a}), and taking the limit $x\rightarrow -\infty$ to eliminate
transient terms,
\[
\la x|p^{+}\ra^{(1)}= \frac{1}{\sqrt{h}}\left[ e^{ipx/\hbar}+
\left(\frac{p-q}{2q}+\frac{mV_{1}}{i\hbar q}\right)
e^{-i qx/\hbar}\right].
\]
To this order, this approach provides a physically meaningless
reflected wave with a momentum smaller than the incident one. 
This indicates that we do not recover in this manner a sensible approximation to the reflectance.
\begin{figure}[top]
\epsfysize=8cm
\centerline{\epsfbox{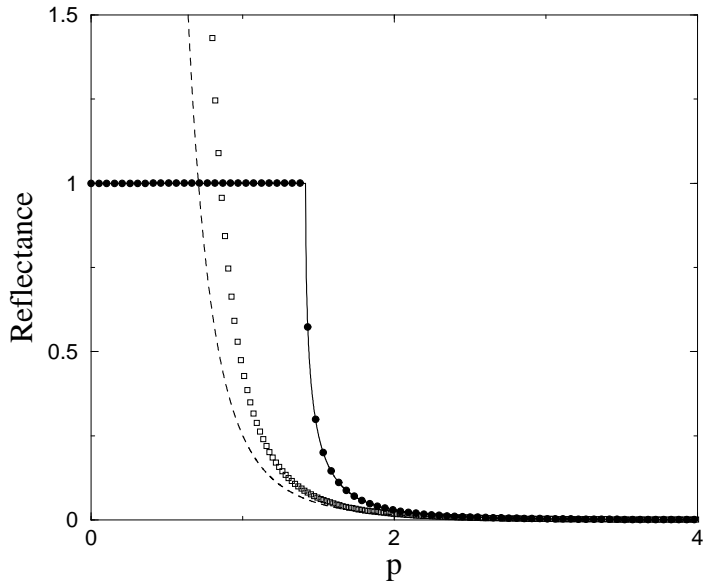}}
\caption{Exact reflectance (solid line), first order Born approximations for the localized 
potential approach (dots) and multichannel method (dashed line), and
second order approximation for the multichannel method (squares) versus $p$.
The potential is given in  (\ref{pote}). $V_0=1\, a.u.$, $V_1=0.01\,
a.u.$, $m=1\, a.u.$}
\end{figure}
Finally, the localized potential (LP) approach of the previous section
gives, to first order in $\wh{V}_s$,
\beq
R^l_{Born-LP}(p)=\frac{p-q}{p+q}-\frac{2\pi miV_1}{p}\la -p_s^-|0\ra
\la 0|p_s^+\ra=\frac{p^2-q^2-4miV_1p/\hbar}{(p+q)^2}.\label{bornlp}
\eeq
The results of  (\ref{bornmm}) and (\ref{bornlp}) are compared in
figure 1, which clearly demonstrate the computational advantage of the
localized potential approach, which starts from much better adapted
initial functions. In particular it is relevant to note that the LP
approach detects the change of regime in the reflectance due to the
energy falling below the asymptotic level, which the multichannel
formalism cannot even suspect in a perturbative scheme. In other
words, the fact that only one channel is open, and  (\ref{uni4})
must hold is overlooked by the perturbative expansion in the MM
scheme, while there is a sharp change in behaviour of the perturbative
expansion in the LP scheme from the one channel to the two channel
case (even though  (\ref{uni4}) does not generically hold if we
restrict ourselves to a finite number of terms).

\section{Discussion}
Given the simplicity of one dimensional step-like potentials, we could
not fail to provide a complete formal scattering theory for
them. However, in pursuing this objective, we have met several
interesting novel aspects with respect to ordinary scattering.  Among
them, the existence of different, all somehow ``natural'',
partitionings of the Hamiltonian is an important one, since it leads
to different formal frameworks.  Working out the details is at the
very least laborious, frequently tedious, and we hope that our compact
presentation and focus on the final results may save some time and
help the practitioners to avoid pitfalls.  With respect to the
three possible methods described, the in/out-formalism has some
elegance, and this was historically our first choice. However, the
zeroth order Hamiltonians, non localized potentials, or Green's
functions are not easy to deal with. This lead us to look for other
possibilities.  Certain manipulations may benefit from the condensed
forms of transmission and reflection amplitudes obtained following the
multichannel method, but in a practical calculation, the localized
potential approach will be generally preferable. It is also the most
economical presentation since it reduces in half the number of
equations needed, and is also the closest to ordinary scattering.

\ack
This work has been supported by Ministerio de 
Ciencia y Tecnolog\'{\i}a (Grants BFM2000-0816-C03-03 and AEN99-0315), The
University of the Basque Country (Grant UPV 063.310-EB187/98), and the
Basque Government (PI-1999-28). A. D. Baute acknowledges an FPI
fellowship by Ministerio de Educaci\'on y Cultura. We thank
M.A. Valle for comments and discussion.

\appendix
\section{Alternative forms of Lippmann-Schwinger equations}
We give an example of the obtention of the alternative LS equations in
(\ref{ppos2}) and (\ref{pneg2}).  Using
\[
\wh{G}_l(z)=\wh{G}_r(z)[1-{V_0}\wh{G}_l(z)]
\]
and $\wh{V}_r=\wh{V}_l-V_0$,
 (\ref{ppos}) for $p>0$ may be written as
\[
|p^+\ra=|p\ra+\wh{G}_r(E_p+i0)\wh{V}_r|p^+\ra
+V_0\wh{G}_l(E_p+i0)[1-\wh{G}_r(E_p+i0)\wh{V}_r]|p^+\ra.
\]
Acting with the operator in parenthesis on $|p^+\ra$, using
$\wh{G}(z)=
\wh{G}_l(z)+\wh{G}_l(z)\wh{V}_l\wh{G}(z)$, 
and (\ref{ppos}), one finds (when operating +i0 must be kept as a
small imaginary number)
\[
|p^+\ra=|p\ra+\wh{G}_r(E_p+i0)\wh{V}_r|p^+\ra+V_0\wh{G}_r(E_p+i0)|p\ra,
\]
but the third term cancels the first one by acting with $\wh{G}_r$ on
$|p\ra$, so that (\ref{ppos2}) (for $p>0$) is obtained.  One may
proceed similarly for the other cases.

\section{Two potential formulae}
In this appendix we shall obtain one of the two potential formulae
used in section VI. The other cases may be obtained similarly.  Assume
that $p<0$. Then,
\beqa
\la p|\wh{V}|\pm p^+\ra&=&\la p|\wh{V}_\theta+\wh{V}_s|\pm p^+\ra\nonumber
\\
&=&\la p|(\wh{V}_\theta+\wh{V}_s)[|\pm p_s^+\ra+
\wh{G}_s(E_p+i0)\wh{T}_s(E_p+i0)|\pm p_s^+\ra]\nonumber
\\
&=&\la p|\wh{V}_\theta|\pm p_s^+\ra +\la p|\wh{V}_\theta
\wh{G}_s(E_p+i0)\wh{T}_s(E_p+i0)|\pm p^+_s\ra+
\la p|\wh{V}_s|\pm p^+\ra\nonumber
\\
&=&\la p|\wh{V}_\theta|\pm p^+_s\ra +\la p_s^-|\wh{V}_s|\pm p^+\ra\,,\nonumber
\eeqa
where we have used  (\ref{lss}) and (\ref{ppos}), the last one
particularized for the pure step potential.


\section*{References}\begin{thebibliography}{10}

\bibitem{AOPRU98}
Aharonov  Y, Oppenheim J, Popescu S, Reznik B and  Unruh W G 1998
{\it Phys. Rev.} A {\bf 57} 4130
\newline (Aharonov  Y, Oppenheim J, Popescu S, Reznik B and  Unruh W G 1997 {\it Preprint} quant-ph/9709031)

\bibitem{BEM01a}
Baute A D, Egusquiza I L and  Muga J G 2001
{\it Phys. Rev.} A , to appear
\newline (Baute A D, Egusquiza I L and  Muga J G 2000 {\it Preprint} quant-ph/0012051)

\bibitem{JW89}
Jaworski W and Wardlaw D M 1989 {\it Phys. Rev.} A {\bf 40} 6210

\bibitem{BF62}
Buslaev V and Fomin V 1962 {\it Vestn. Leningr. Univ. Se. 4: Fiz. Khim} {\bf 17} 56

\bibitem{CK85}
Cohen A and Kappeler T 1985 {\it Indiana Univ. Math. J.} {\bf 34} 127 

\bibitem{Legendre82}
Legendre J 1982  PhD thesis, Acad. Montpellier, Univ. Sci. Tech. Languedoc

\bibitem{Aktosun99}
Aktosun T 1999 {\it J. Math. Phys.} {\bf 40} 5289

\bibitem{BEM01}
Baute A D, Egusquiza I L and  Muga J G 2001 {\it Int. J. Theor. Phys.}, to appear
\newline (Baute A D, Egusquiza I L and  Muga J G 2000 {\it Preprint} quant-ph/0007079)

\bibitem{HWZ77}
Hammer C L, Weber T A  and Zidell V S 1977 {\it Am. J. Phys.} {\bf 45} 933

\bibitem{Aguiar93}
de~Aguiar M A M 1993 {\it Phys. Rev. A} {\bf 48} 2567

\end{thebibliography}
\end{document}